\documentclass[preprint,authoryear,12pt]{elsarticle}
\usepackage{graphics,epsfig}
\usepackage{graphicx}
\usepackage{amssymb,amstext,amsmath}

\usepackage{booktabs}
\usepackage{natbib}
\usepackage[latin1]{inputenc}
\usepackage{epstopdf}
\usepackage{mathtools}
\begin{document}
\newcommand{\balpha}{\boldsymbol{\alpha}}
\newcommand{\bbeta}{\boldsymbol{\beta}}
\newcommand{\bgamma}{\boldsymbol{\gamma}}
\newcommand{\bdelta}{\boldsymbol{\delta}}
\newcommand{\bepsilon}{\boldsymbol{\epsilon}}
\newcommand{\bvarepsilon}{\boldsymbol{\varepsilon}}
\newcommand{\bzeta}{\boldsymbol{\zeta}}
\newcommand{\bfoldeta}{\boldsymbol{\eta}}
\newcommand{\btheta}{\boldsymbol{\theta}}
\newcommand{\bvartheta}{\boldsymbol{\vartheta}}
\newcommand{\biota}{\boldsymbol{\iota}}
\newcommand{\bkappa}{\boldsymbol{\kappa}}
\newcommand{\blambda}{\boldsymbol{\lambda}}
\newcommand{\bmu}{\boldsymbol{\mu}}
\newcommand{\bnu}{\boldsymbol{\nu}}
\newcommand{\bxi}{\boldsymbol{\xi}}
\newcommand{\bpi}{\boldsymbol{\pi}}
\newcommand{\bvarpi}{\boldsymbol{\varpi}}
\newcommand{\brho}{\boldsymbol{\rho}}
\newcommand{\bvarrho}{\boldsymbol{\varrho}}
\newcommand{\bsigma}{\boldsymbol{\sigma}}
\newcommand{\bvarsigma}{\boldsymbol{\varsigma}}
\newcommand{\btau}{\boldsymbol{\tau}}
\newcommand{\bupsilon}{\boldsymbol{\upsilon}}
\newcommand{\bphi}{\boldsymbol{\phi}}
\newcommand{\bvarphi}{\boldsymbol{\varphi}}
\newcommand{\bchi}{\boldsymbol{\chi}}
\newcommand{\bpsi}{\boldsymbol{\psi}}
\newcommand{\bomega}{\boldsymbol{\omega}}
\newcommand{\bGamma}{\boldsymbol{\Gamma}}
\newcommand{\bDelta}{\boldsymbol{\Delta}}
\newcommand{\bTheta}{\boldsymbol{\Theta}}
\newcommand{\bLambda}{\boldsymbol{\Lambda}}
\newcommand{\bXi}{\boldsymbol{\Xi}}
\newcommand{\bPi}{\boldsymbol{\Pi}}
\newcommand{\bSigma}{\boldsymbol{\Sigma}}
\newcommand{\bUpsilon}{\boldsymbol{\Upsilon}}
\newcommand{\bPhi}{\boldsymbol{\Phi}}
\newcommand{\bPsi}{\boldsymbol{\Psi}}
\newcommand{\bOmega}{\boldsymbol{\Omega}}
\newcommand{\llbracket}{[\![}
\newcommand{\rrbracket}{]\!]}

\begin{frontmatter}
\title{Thermodynamic dislocation theory for non-uniform plastic deformations}
\author{K. C. Le}
\address{Lehrstuhl f\"{u}r Mechanik - Materialtheorie, Ruhr-Universit\"{a}t Bochum, D-44780 Bochum, Germany}
\begin{abstract}	
The present paper extends the thermodynamic dislocation theory developed by Langer, Bouchbinder, and Lookman to non-uniform plastic deformations. The free energy density as well as the positive definite dissipation function are proposed. The governing equations are derived from the variational equation. As illustration, the problem of plane strain constrained shear of single crystal deforming in single slip is solved within the proposed theory. 
\end{abstract}
\begin{keyword}
dislocations \sep thermodynamics \sep configurational temperature \sep plastic yielding \sep strain rate.
\end{keyword}

\end{frontmatter}

\section{Introduction}
\label{intro}

Macroscopically observable plastic deformations of single crystals and polycrystalline materials are caused by nucleation, multiplication and motion of dislocations. The interesting question arrises in this connection: is entropy of dislocations relevant to thermodynamics of plasticity and should we take it into account in the continuum dislocation theory? At first glance, since the energy of a single dislocation is so large that ordinary thermal fluctuations cannot create or annihilate it, the kinetic-vibrational temperature of the crystal seems to be  irrelevant to the nucleation and motion of dislocations. On the other hand, as the entropy of disorder induced by a dislocation is extremely small compared with the total entropy of the crystal, the standard thermodynamics of crystal plasticity (see, e.g., \citep{Rice1971,Halphen1975,Lubliner2008} and the references therein) have been ignoring the entropy of dislocations completely. However, the novel approach proposed by \citet{Langer2010}, termed LBL-theory for short, (see also \citep{Langer2015,Langer2016,Langer2017}) has shown that the entropy of a large number of dislocations is an essential ingredient of a theory of dislocation-mediated plastic flow. The crucial step done in these papers is to decouple the system of crystal containing dislocations into configurational and kinetic-vibrational subsystems. The configurational degrees of freedom describe the relatively slow, i.e. infrequent, atomic rearrangements that are associated with the irreversible movement of dislocations; the kinetic-vibrational degrees of freedom describe the fast vibrations of atoms in the lattice. The governing equations of LBL-theory are based on the kinetics of thermally activated dislocation depinning and irreversible thermodynamics of driven systems. This LBL-theory has been successfully used to simulate the stress-strain curve for copper over fifteen decades of strain rate, and for temperatures between room temperature and about one third of the melting temperature. Only one fitting parameter is required to get the full agreement with the experiment conducted by \citet{Follansbee1998} over a wide range of temperatures and strain rates. The theory has been extended to include the interaction between two subsystems by \citet{Langer2017} and used to simulate the stress-strain curves for aluminum and steel which exhibit the thermal softening \citep{Le2017} in full agreement with the experiments conducted in \citep{Shi1997,Abbot2007}. 

The LBL-theory as well as its extensions apply to the uniform plastic deformations, where the dislocations are neutral in the sense that their resultant Burgers vector vanishes. \citet{Ashby1970} called this sort of dislocations statistically stored, but we prefer the shorter and more precise name of redundant dislocations given earlier by \citet{Cottrell1964}. When dealing with non-uniform plastic deformations as, say, the torsion of a bar, the bending of a beam, or deformations of two-phase alloys or polycrystals, another sort of dislocations appear in addition to redundant dislocations to accommodate the plastic deformation gradient and to guarantee the compatibility of the total deformation \citep{Nye1953,Bilby1955,Kroener1955,Ashby1970}. Most of the contemporary experts in dislocation theory accept the suggestion of \citet{Ashby1970} to call these dislocations ``geometrically necessary''. However, from the point of view of statistical mechanics of dislocations (see, e.g., \citep{Berdichevsky2006b,Limkumnerd2008,Poh2013,Zaiser2015}) the term excess dislocations seems more precise. Note that, in recent years, the density of excess dislocations can be indirectly measured by the high resolution electron backscatter diffraction technique (EBSD) \citep{Kysar2010}. Although the percentage of excess dislocations in severely and plastically deformed crystals is small, they play the prominent role in the formation of microstructure \citep{Ortiz1999,Huang2001,Kochmann2009,Le2012,Koster2015} and the size effects \citep{Fleck1994,Nix1998,Hansen2004,Kaluza2011,Le2013}. The continuum dislocation theory (CDT) incorporating the density of excess dislocations, as developed by \citet{Berdichevsky1967,Le1996,Gurtin2002,Berdichevsky2006a,Le2014}, can indeed capture the formation of microstructure and the size effects. However, the main drawback of this strain-gradient plasticity approach is the absence of redundant dislocations and the configurational entropy (or its dual quantity, the configurational temperature). Since these two quantities are crucial for the isotropic work hardening  \citep{Langer2010}, the continuum dislocation theory needs an essential revision.

The aim of this paper is to provide the revision of CDT. By including its two missing quantities, configurational entropy (or its dual quantity, configurational temperature) and density of redundant dislocations, as state variables in the constitutive equations, we get the synthesis of CDT and LBL-theory that is a truly dynamical theory valid for arbitrary non-uniform plastic deformations. We call it thermodynamic dislocation theory (TDT). The latter reduces to the LBL-theory for uniform deformations where the density of excess dislocations vanishes. As an illustration, the problem of plane strain constrained shear of single crystal deforming in single slip is solved within the proposed theory. It will be shown that this problem can approximately be reduced to  two separate problems: (i) find the flow stress, the density of dislocations, and the configurational temperature according to LBL-theory, (ii) find the stress, strain, and density of excess dislocations from the relaxed energy minimization. The second problem is solved by the method developed in \citep{Le2008}. The thickness of the boundary layers where excess dislocations pile up depends on the flow stress obtained from the solution of the first problem. The stress-strain curve is shown to be temperature- and rate-sensitive, and exhibits also the size effect.  

The paper is organized as follows. In Section 2 the kinematics of TDT is laid down. Section 3 proposes the thermodynamic framework for this TDT. In Section 4 the problem of anti-plane constrained shear is analyzed. Section 5 presents the numerical solution of this problem and discusses the distribution of excess dislocations, the stress-strain curve, and the size effect. Finally, Section 6 concludes the paper.  

\section{Kinematics}
\label{sec:1}
In the following we restrict ourselves to the plane strain deformation of a single crystal having only one active slip system. For simplicity we shall use some fixed rectangular cartesian coordinates and denote by $\mathbf{x}=(x_1,x_2)$ the position vector of a generic material point of the crystal's cross section. Kinematical quantities characterizing the observable deformation of this single crystal are the displacement field $\mathbf{u}(\mathbf{x},t)$ and the plastic distortion field $\bbeta  (\mathbf{x},t)$ that is in general incompatible. For single crystals having one active slip system, the plastic distortion is given by
\begin{equation}
\label{eq:1.1}
\bbeta  (\mathbf{x},t)=\beta (\mathbf{x},t) \mathbf{s}\otimes \mathbf{m} \quad (\beta _{ij}= \beta s_im_j),
\end{equation}  
with $\beta (\mathbf{x},t)$ being the plastic slip, where the pair of constant and mutually orthogonal unit vectors $\mathbf{s}$ and $\mathbf{m}$ lye in the $(x_1,x_2)$-plane and denote the slip direction and the normal to the slip planes, respectively. The edge dislocations causing this plastic slip have dislocation lines parallel to the $x_3$-axis and the Burgers vector parallel to $\mathbf{s}$. Thus, there are altogether three degrees of freedom at each point of this generalized continuum. Later on, two additional internal variables will be introduced. In this Section, equivalent formulas for the components of tensors are also given on the same line in brackets, where the Latin lower indices running from 1 to 2 indicate the projections onto the corresponding coordinates. We use Einstein's summation convention, according to which summation over repeated Latin indices from 1 to 2 is understood. We see immediately from \eqref{eq:1.1} that $\text{tr}\bbeta  =\beta _{ii}=0$, so the plastic distortion is volume preserving. 

\begin{figure}[htb]
\centering \includegraphics[height=7cm]{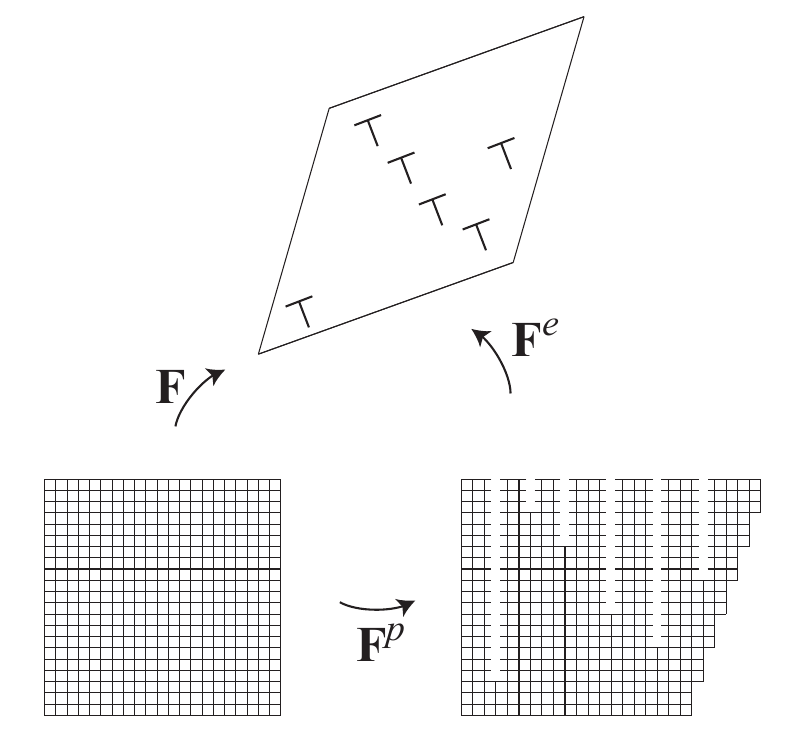} 
\caption{Additive decomposition}
\label{fig:Nhslip}
\end{figure}

The total compatible strain tensor field can be obtained from the displacement field according to
\begin{equation*}
\bvarepsilon =\frac{1}{2}(\mathbf{u}\nabla +\nabla \mathbf{u}) \quad (\varepsilon _{ij}=\frac{1}{2}(u_{i,j}+u_{j,i})).
\end{equation*}
The incompatible plastic strain tensor field is the symmetric part of the plastic distortion field
\begin{equation*}
\bvarepsilon ^p =\frac{1}{2}(\bbeta  +\bbeta  ^T) \quad (\varepsilon ^p_{ij}=\frac{1}{2}(\beta_{ij}+\beta_{ji})).
\end{equation*}
For small strains we shall use the additive decomposition of the total strain into the elastic and plastic parts. Therefore, the elastic strain tensor field is equal to
\begin{equation*}
\bvarepsilon ^e=\bvarepsilon -\bvarepsilon ^p \quad (\varepsilon ^e_{ij}=\varepsilon _{ij}-\varepsilon ^p_{ij}).
\end{equation*}
Although not absolutely necessary, we introduce for the illustration purpose the elastic distortion tensor field according to
\begin{equation*}
\bbeta  ^e=\mathbf{u}\nabla -\bbeta  \quad (\beta ^e_{ij}=u_{i,j}-\beta _{ij}).
\end{equation*}
The relationship between these three distortion fields is illustrated schematically in Fig.~\ref{fig:Nhslip}, where $\mathbf{F}=\mathbf{I}+\mathbf{u}\nabla $, $\mathbf{F}^p=\mathbf{I}+\bbeta  $, $\mathbf{F}^e=\mathbf{I}+\bbeta  ^e$. Looking at this Figure we see that the non-uniform plastic distortion $\bbeta  $ is that {\it creating} excess dislocations (either inside or at the boundary of the volume element) or {\it changing} their positions in the crystal without deforming the crystal lattice. In contrary, the elastic distortion $\bbeta ^e$ deforms the crystal lattice having {\it frozen} dislocations \citep{Le2014}.

Under the plane strain deformation the tensor of excess dislocation density, introduced by \citet{Nye1953,Bilby1955,Kroener1955}, $\balpha=-\bbeta \times \nabla $, has only two non-zero components $\alpha _{i3}=s_i (\partial _s \beta )$, where $\partial _s=s_i\partial _i$ denotes the derivative in the direction $\mathbf{s}$. Thus, the resultant Burgers vector of all excess dislocations, whose lines cross the infinitesimal area $\mathrm{d}a$ perpendicular to the $x_3$-axis is
\begin{displaymath}
\mathrm{d}B_i=\alpha _{i3}\mathrm{d}a=s_i (\partial _s \beta )\mathrm{d}a.
\end{displaymath}
Therefore, the scalar density of excess dislocations is characterized by 
\begin{equation}\label{excess}
\rho ^g=\frac{|\mathrm{d}B|}{b\mathrm{d}a}=\frac{1}{b}|\partial _s \beta |,
\end{equation}
where $b$ is the magnitude of the Burgers vector. This density can be indirectly measured by the high resolution electron backscatter diffraction technique (EBSD) \citep{Kysar2010}. In addition to the excess dislocations there exist another family of dislocations which cannot be obtained from \eqref{excess} but nevertheless may have significant influences on the nucleation of excess dislocations and the work hardening of crystals. For any closed circuit surrounding an infinitesimal area (in the sense of continuum mechanics) the resultant Burgers vector of these dislocations always vanishes, so the closure failure caused by the incompatible plastic distortion is not affected by them. Following \citet{Cottrell1964} and \citet{Weertman1996} we call these dislocations redundant. As a rule, the redundant dislocations in unloaded crystals at low temperatures exist in form of dislocation dipoles. The simple reason for this is that the energy of a dislocation dipole is much smaller than that of dislocations apart, so this bounded state of dislocations of opposite sign renders low energy to the whole crystal. From the other side, due to their low energy, the dislocation dipoles can be created by the mutual trapping of dislocations of different signs in a random way or, eventually, by thermal fluctuations. Let us denote the density of redundant dislocations by $\rho ^r$. The total dislocation density is thus
\begin{equation*}
\rho =\rho ^g+\rho ^r.
\end{equation*}

\section{Thermodynamic dislocation theory}
\label{sec:2}

To set up phenomenological models of crystals with continuously distributed dislocations using the methods of non-equilibrium thermodynamics of driven system let us begin with the free energy density. As a function of the state, the free energy density may depend only on the state variables. Following \citet{Kroener1992,Langer2016} we will assume that the elastic strain $\bvarepsilon^e$, the dislocation densities $\rho ^r$ and $\rho ^g$, the kinetic-vibrational temperature $T$, and the configurational temperature $\chi $ characterize the current state of the crystal, so these quantities are the state variables of the continuum dislocation theory. The reason why the plastic distortion $\bbeta  $ cannot be qualified for the state variable is that it depends on the cut surfaces and consequently on the whole  history of creating dislocations (for instance, climb or glide dislocations are created quite differently). Likewise, the gradient of plastic strain tensor $\bvarepsilon^p$ cannot be used as the state variable by the same reason. In contrary, the dislocation densities $\rho ^r$ and $\rho ^g$ depend only on the characteristics of  dislocations in the current state (Burgers vector and positions of dislocation lines) and not on how they are created, so they are the proper state variable. We restrict ourself to the isothermal processes, so the kinetic-vibrational temperature $T$ is assumed to be constant and can be dropped in the list of arguments of the free energy density. Two state variables, $\bvarepsilon^e$ and $\rho ^g$ are dependent variables as they are expressible through the degrees of freedom $\mathbf{u}$ and $\beta$, while two others, $\rho ^r$ and $\chi $, can be regarded as independent internal variables. Our main assumption for the free energy density is
\begin{equation}\label{eq:2.1}
\psi = \frac{1}{2}\lambda (\varepsilon ^e_{kk})^2+\mu \varepsilon ^e_{ij}\varepsilon ^e_{ij}+\gamma_D\rho ^r+\frac{\gamma_D}{a^2}\ln \frac{1}{1-a^2\rho ^g} -\chi (- \rho \ln (a^2 \rho )+\rho )/L.
\end{equation}
The two first terms in \eqref{eq:2.1} describe energy of crystal due to the elastic strain. The third term is the self-energy of redundant dislocations, with $\gamma_D$ being the energy of one dislocation per unit length. The fourth term is the energy of excess dislocations. This logarithmic energy term stems from two facts: (i) energy of excess dislocations for small dislocation densities must be $\gamma_D\rho ^g$ like that of redundant dislocations, and (ii) there exists a saturate state of maximum disorder and infinite configurational temperature characterized by the admissibly closest distance between excess dislocations, $a$. The logarithmic term \citep{Berdichevsky2006b} ensures a linear increase of the energy for small dislocation density $\rho ^g$ and tends to infinity as $\rho ^g$ approaches the saturated dislocation density $1/a^2$ hence providing an energetic barrier against over-saturation. The last term has been introduced by \citet{Langer2016}, with $S_C=(-\rho \ln (a^2 \rho )+\rho )/L$ being the configurational entropy of dislocations.

With this free energy density we can now write down the energy functional of the crystal. Let the cross section of the undeformed single crystal occupy the region $\mathcal{A}$ of the $(x_1,x_2)$-plane. The boundary of this region, $\partial \mathcal{A}$, is assumed to be the closure of union of two non-intersecting curves, $\partial _k$ and $\partial _s$. Let the displacement vector $\mathbf{u}(\mathbf{x},t)$ be a given smooth function of coordinates (clamped boundary), and, consequently, the plastic slips $\beta (\mathbf{x},t)$ vanish 
\begin{equation}\label{eq:2.2}
\mathbf{u}(\mathbf{x},t)=\tilde{\mathbf{u}}(\mathbf{x}),\quad \beta (\mathbf{x},t)=0 \quad \text{for $\mathbf{x}\in \partial _k$}.
\end{equation}
The remaining part $\partial _s$ of the boundary is assumed to be traction-free. If no body force acts on this crystal, then its energy functional per unit depth is defined as
\begin{equation*}
I[\mathbf{u}(\mathbf{x},t),\beta (\mathbf{x},t),\rho ^r(\mathbf{x},t),\chi (\mathbf{x},t))]=\int_{\mathcal{A}}\psi (\bvarepsilon^e,\rho ^r, \rho ^g,\chi )\,\mathrm{d}a,
\end{equation*}
with $\mathrm{d}a=\mathrm{d}x_1\mathrm{d}x_2$ denoting the area element. 

Under the increasing load the resolved shear stress also increases, and when it reaches the Taylor stress, dislocations dipoles dissolve and begin to move until they are trapped again by dislocations of opposite sign. During this motion dislocations always experience the resistance causing the energy dissipation. The increase of dislocation density as well as the increase of configurational temperature also lead to the energy dissipation. Neglecting the dissipation due to internal viscosity associated with the strain rate, we propose the dissipation potential in the form 
\begin{equation}\label{eq:2.4}
D(\dot{\beta},\dot{\rho},\dot{\chi})=\tau _Y \dot{\beta }+\frac{1}{2}d_\rho \dot{\rho }^2+\frac{1}{2}d_\chi \dot{\chi }^2,
\end{equation}
where $\tau_Y$ is the flow stress during plastic yielding, $d_\rho $ and $d_\chi $ are still unknown functions, to be determined later. The first term in \eqref{eq:2.4} is the plastic power which is assumed to be homogeneous function of first order with respect to the plastic slip rate \citep{Puglisi2005}. The other two terms describe the dissipation caused by the multiplication of dislocations and the increase of configurational temperature \citep{Langer2010}. Based on Hooke's law, Orowan's equation and the kinetics of dislocation depinning \citep{Langer2010}, the following equation holds true for $\tau _Y$
\begin{equation}
\label{eq:2.5}
\dot{\tau }_Y=\mu \frac{q_0}{t_0}\left[ 1-\frac{q(\tau_Y,\rho)}{q_0}\right] .
\end{equation}
In Eq.~\eqref{eq:2.5} $q_0/t_0=\dot{\gamma }^r=2\dot{\varepsilon}_{ij}s_im_j$ is the rate of the resolved shear strain, assumed here to be positive, with $t_0=10^{-12}$s being the characteristic microscopic time scale. The rate of plastic slip is $\dot{\beta}=q(\tau_Y,\rho)/t_0$, where
\begin{equation*}
q(\tau_Y,\rho)=b\sqrt{\rho }\exp \left[ -\frac{1}{\theta }e^{-\tau _Y/\tau _T}\right] .
\end{equation*}
In this equation the dimensionless temperature is introduced as $\theta =T/T_P$, with $T_P$ being the pinning energy barrier, and $\tau _T=\mu _T b\sqrt{\rho }$ the Taylor stress. Since the dislocation mediated plastic flow is the irreversible process, we derive the governing equations from the following variational formulation: the true displacement field $\check{\mathbf{u}}(\mathbf{x},t)$, the true plastic slips $\check{\beta }(\mathbf{x},t)$, the true density of redundant dislocations $\check{\rho }^r(\mathbf{x},t)$, and the true configurational temperature $\check{\chi }(\mathbf{x},t)$ obey the variational equation
\begin{equation}
\label{eq:2.8}
\delta I+\int_{\mathcal{A}} \left( \frac{\partial D}{\partial \dot{\beta }}\delta \beta +\frac{\partial D}{\partial \dot{\rho }}\delta \rho +\frac{\partial D}{\partial \dot{\chi }}\delta \chi \, \right) \,\mathrm{d}a=0
\end{equation}
for all variations of admissible fields $\mathbf{u}(\mathbf{x},t)$, $\beta (\mathbf{x},t)$, $\rho ^r(\mathbf{x},t)$, and $\chi (\mathbf{x},t)$ satisfying the constraints \eqref{eq:2.2}. 

Varying the energy functional with respect to $u_{i}$ we obtain the quasi-static equation of equilibrium of macro-forces
\begin{equation}
\label{eq:2.9}
\sigma _{ij,j}=0, \quad \sigma _{ij}=\frac{\partial \psi}{\partial \varepsilon ^e_{ij}}=\lambda \varepsilon ^e_{kk}\delta _{ij}+2\mu \varepsilon ^e_{ij}, 
\end{equation}
which are subjected to the boundary conditions \eqref{eq:2.2} and 
\begin{equation*}
\sigma _{ij}n_j=0 \quad \text{on $\partial _s$}.
\end{equation*}
Taking the variation of $I$ with respect to three other quantities $\beta $, $\rho ^r$, and $\chi $ and requiring that Eq.~\eqref{eq:2.8} is satisfied for their admissible variations, we get three equations
\begin{equation}
\label{eq:2.10}
\begin{split}
\tau -\tau_B -\tau _Y=0, 
\\
(e_D+\chi \ln (a^2\rho))/L +d_\rho \dot{\rho }=0, 
\\
(\rho \ln (a^2\rho) -\rho)/L+d_\chi \dot{\chi }=0,
\end{split}
\end{equation}
where $\tau =\sigma _{ij}s_im_j$ is the resolved shear stress (Schmid stress), while $\tau_B=-\partial^2\psi /\partial (\rho ^g)^2 \beta_{,ss}$ the back stress. The first equation of \eqref{eq:2.10}, valid under the condition $\dot{\beta}>0$, can be interpreted as the balance of microforces acting on dislocations. This equation is subjected to the Dirichlet boundary condition \eqref{eq:2.2}$_2$ on $\partial _k$ and
\begin{displaymath}
\frac{\partial \psi}{\partial \rho^g } =\gamma_D \quad \text{on $\partial _s$}.
\end{displaymath}
Combined with \eqref{eq:2.5} and \eqref{eq:2.9}, \eqref{eq:2.10} yield the equations of motion of dislocations. Note that, for the fast loading when the inertia term becomes essential, Eq.~\eqref{eq:2.9} should be modified to
\begin{equation*}
\varrho \ddot{u}_i=\sigma _{ij,j},  
\end{equation*}
with $\varrho $ being the mass density.

We require that, for the uniform total and plastic deformations, system \eqref{eq:2.5}, \eqref{eq:2.9}, \eqref{eq:2.10} reduces to the system of equations of LBL-theory \citep{Langer2010}
\begin{align}
\dot{\tau } & =  \mu \frac{q_0}{t_0}\left[1-\frac{q(\tau ,\rho )}{q_0}\right] ,  \notag
\\
\dot{\chi } & =  \mathcal{K} \tau \frac{q(\tau ,\rho )}{t_0} \left[ 1-\frac{\chi }{\chi ^{ss}(q)} \right], \label{eq:2.11}
\\
\dot{\rho } & = \mathcal{K}_\rho \frac{\tau }{a^2\nu (T,\rho ,q_0)^2}\frac{q(\tau ,\rho )}{t_0}\left[ 1-\frac{\rho }{\rho ^{ss}(\chi )} \right] . \notag
\end{align}
As compared to the original equations derived in \citep{Langer2010} there are some changes in notations to make them consistent with those employed in this paper: the resolved shear stress is denoted by $\tau $ instead of $\sigma $, the resolved shear strain rate by $\dot{\gamma }^r$ instead of $\dot{\epsilon }$, while the plastic slip rate by $\dot{\beta }$ instead of $\dot{\varepsilon}^p$. The steady-state configurational temperature is denote by $\chi ^{ss}$, while the steady-state dislocation density is
\begin{displaymath}
\rho ^{ss}(\chi )=\frac{1}{a^2}e^{-e_D/\chi }.
\end{displaymath}
Finally, $\nu (T,\rho ,q_0)$ is defined as follows
\begin{displaymath}
\nu (T,\rho ,q_0)=\ln \left( \frac{T_P}{T}\right) - \ln \left[ \frac{1}{2}\ln \left( \frac{b^2\rho }{q_0^2}\right) \right] .
\end{displaymath}
Since the total and plastic deformation are uniform, Eq.~\eqref{eq:2.9} is satisfied identically. By the same reason, the second and fourth terms in \eqref{eq:2.10}$_1$ disappear, so $\tau =\tau _Y$, and in combination with Eq.~\eqref{eq:2.5}, this leads to \eqref{eq:2.11}$_1$. Two remaining equations of \eqref{eq:2.10}$_{2,3}$ reduce to \eqref{eq:2.11}$_{2,3}$  if we choose
\begin{align}
\label{eq:2.12}
d_\chi &=\frac{\rho -\rho \ln (a^2\rho) }{L\mathcal{K} \tau_Y \frac{q(\tau_Y ,\rho )}{t_0} \left[ 1-\frac{\chi }{\chi ^{ss}(q)} \right] },
\\
d_\rho &=\frac{-e_D-\chi \ln (a^2\rho)}{L\mathcal{K}_\rho \frac{\tau_Y }{a^2\nu (T,\rho ,q_0)^2}\frac{q(\tau_Y ,\rho )}{t_0}\left[ 1-\frac{\rho }{\rho ^{ss}(\chi )} \right]} . \label{eq:2.13}
\end{align}
Note that, for $\rho $ changing between 0 and $\rho ^{ss}<1/a^2$, both numerators on the right-hand sides of \eqref{eq:2.12} and \eqref{eq:2.13} are positive, and the dissipative potential \eqref{eq:2.4} is positive definite as required by the second law of thermodynamics.

\section{Plane strain constrained shear}
\label{sec:3}

\begin{figure}[htb]
\centering \includegraphics[height=7cm]{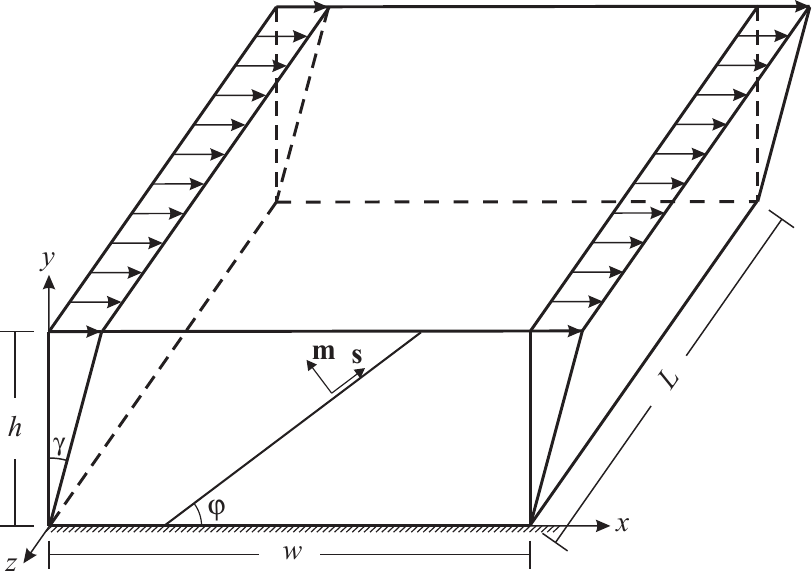}
\caption{Plane constrained shear}
\label{fig:planeshear}
\end{figure}

As an application of the proposed theory let us consider the single crystal layer having a rectangular cross-section of width $w$ and height $h$, $0 \leq x \leq w$, $0 \leq y \leq h$ and undergoing a plane strain constrained shear deformation (see Fig.~\ref{fig:planeshear}). The single crystal is placed in a hard device, which models the grain boundary, with prescribed displacements at its upper and lower sides as
\begin{equation}\label{SS_boundary}
    u(0,t)=0, \quad v(0,t)=0, \quad u(h,t)=\gamma(t) h, \quad v(h,t)=0,
\end{equation}
where $u(y,t)$ and $v(y,t)$ are the longitudinal and transverse displacements, respectively, with $\gamma(t)$ being the overall shear strain. We assume that the length of the strip $L$ is very large, and the width $w$ is much greater than the height $h$ ($L \gg w \gg h$) to neglect the end effects near the boundaries $x=0$ and $x=w$ and to have the stresses and strains depending only on one variable $y$ in the central part of the strip (due to the almost translational invariance in $x$-direction). We admit only one active slip system, with the slip directions inclined at an angle $\varphi $ to the $x$-axis and the edge dislocations whose lines are parallel to the $z$-axis. We aim at determining the densities of total and excess dislocations, configurational temperature, and the stress-strain curve as function of $\gamma (t)$ by the thermodynamic  dislocation theory proposed in the previous Section. 

For the plane strain state the in-plane components of the strain tensor, $\varepsilon_{ij} = \frac{1}{2} (u_{i,j} + u_{j,i})$, are given by
\begin{equation} \label{SS_totalstrain}
\varepsilon _{xx}=0,\quad \varepsilon_{xy}=\varepsilon_{yx}=\frac{1}{2}u_{,y},\quad \varepsilon_{yy}=v_{,y}.
\end{equation}
The active slip system inclined at the angle $\varphi $ to the $x$-axis has the vectors  $\mathbf{s}=(\cos \varphi , \sin \varphi ,0)$ and $\mathbf{m}=(-\sin \varphi , \cos \varphi ,0)$, therefore the plastic distortion is
\begin{equation}
 \beta_{i j}=
    \begin{pmatrix}
        -\beta \sin \varphi \cos \varphi & \beta \cos ^{2} \varphi  \\
        -\beta \sin ^{2} \varphi  & \beta \sin \varphi \cos \varphi  
        \end{pmatrix}. \label{SS_Betaij}
\end{equation}
Since the edge dislocations cannot reach the grain boundaries, we require that
\begin{equation*}
\beta (0,t)=\beta (h,t)=0.
\end{equation*}
It follows from \eqref {SS_Betaij}  that the non-zero components of the plastic strain tensor $\varepsilon^{p}_{ij}=\frac{1}{2}(\beta_{i j}+\beta_{j i})$ are
\begin{equation}\label{SS_plasticstrain}
\varepsilon^{p}_{xx}=-\frac{1}{2}\beta \sin 2\varphi ,\quad \varepsilon^{p}_{xy}=\varepsilon^{p}_{yx}=\frac{1}{2}\beta \cos 2 \varphi ,\quad
\varepsilon^{p}_{yy}=\frac{1}{2}\beta \sin 2 \varphi.
\end{equation}
With \eqref{SS_totalstrain} and \eqref{SS_plasticstrain}, the
non-zero components of the elastic strain tensor, $\varepsilon ^e_{ij}=\varepsilon _{ij}-\varepsilon ^p_{ij}$, read
\begin{equation*}
	\varepsilon^{e}_{xx}=\frac{1}{2}\beta \sin 2\varphi ,\quad
	\varepsilon^{e}_{xy}=\varepsilon^{e}_{yx}=\frac{1}{2}(u_{,y}-\beta \cos 2 \varphi ),\quad  \varepsilon^{e}_{yy}=v_{,y}-\frac{1}{2}\beta \sin 2\varphi.
\end{equation*}
The scalar density of excess dislocations (or the number of excess dislocations per unit area) equals
\begin{equation*}
    \rho^g =\frac{1}{b}|\beta_{,y}||\sin \varphi|. 
\end{equation*}

Under the assumptions made, the energy functional becomes 
\begin{multline}
I[u,v,\beta,\rho ^r,\chi ]=w L \int_{0}^{h} \biggl[\frac{1}{2}\lambda v_{,y}^{2} + \frac{1}{2}\mu ( u_{,y}-\beta \cos 2\varphi
    )^{2} + \frac{1}{4}\mu \beta^{2} \sin^{2} 2\varphi \\
     + \mu (v_{,y}-\frac{1}{2}\beta \sin
    2\varphi )^{2}+\gamma_D\rho ^r + \frac{\gamma_D}{a^2} \ln \frac{1}{ 1-\frac{a^2}{b}|\beta_{,y}||\sin \varphi|
    }-\chi (- \rho \ln (a^2 \rho )+\rho )/L \biggr] \, \mathrm{d} y. \label{SS_energyfunctional}
\end{multline}
The energy functional \eqref{SS_energyfunctional} can be reduced to a functional depending on $\beta$, $\rho ^r$, and $\chi $ only. Indeed, taking the variation of \eqref{SS_energyfunctional} with respect to $u$ and $v$, we derive the equilibrium equations
\begin{equation*}
\begin{split}
    \mu (u_{,yy} - \beta_{,y} \cos 2\varphi )=0,
    \\
    \mu (\frac{1}{\kappa} v_{,yy} -  \beta_{,y} \sin 2\varphi  ) =0,
\end{split}
\end{equation*}
subjected to the boundary conditions \eqref{SS_boundary}, with $\kappa = \frac{\mu}{\lambda + 2 \mu}$. Integrating these equations and taking conditions \eqref{SS_boundary} into account, we obtain
\begin{align}
     u_{,y}&= \gamma + (\beta- \langle \beta \rangle) \cos 2\varphi. \label{SS_uy}
\\
     v_{,y}&= \kappa (\beta - \langle \beta \rangle) \sin 2\varphi, 
\end{align}
where $\langle \beta \rangle =\frac{1}{h}\int_0^h \beta \,dy$. With these results the energy functional reduces to
\begin{multline}
I=w L \int_0^h \biggl[ \frac{1}{2} \mu \kappa \langle \beta \rangle ^2 \sin ^2  2\varphi +  \frac{1}{2}\mu (\langle \beta \rangle \cos 2\varphi  - \gamma)^2 +  \frac{1}{2}\mu (1-\kappa) \beta ^2 \sin ^2 2\varphi \\
    +\gamma_D\rho ^r + \frac{\gamma_D}{a^2} \ln \frac{1}{ 1-\frac{a^2}{b}|\beta_{,y}||\sin \varphi|
    }-\chi (- \rho \ln (a^2 \rho )+\rho )/L \biggr]\,    \mathrm{d} y. \label{SS_energyFinal}
\end{multline}

As will be seen later, under the resolved shear stress dislocations of different sign move along the slip lines in the opposite directions. Since dislocations cannot reach the grain boundaries, the excess dislocations will pile up against the upper and lower boundaries leaving the central part of the layer free of them. Since the thickness of the boundary layers where excess dislocations pile up is quite small compared to $h$, the plastic slip $\beta $ is nearly uniform, and according to \eqref{SS_uy} the rate of resolved shear strain is almost everywhere constant and equal to $\dot{\gamma }\cos ^2 \varphi$. Neglecting a small difference in the resolved shear strain rate in the boundary layers, we reduce the determination of $\tau_Y$, $\rho $, and $\chi $ in the first approximation to the solution of \eqref{eq:2.11}, with $\tau $ being replaced by $\tau _Y$ and $q_0=t_0 \dot{\gamma }\cos ^2 \varphi$. After knowing $\tau _Y$, $\rho $, and $\chi $, the variational equation \eqref{eq:2.8} reduces to minimizing the following ``relaxed'' energy functional
\begin{multline}
I_d = w L \int_0^h \biggl[ \frac{1}{2} \mu \kappa \langle \beta \rangle ^2 \sin ^2  2\varphi +  \frac{1}{2}\mu (\langle \beta \rangle \cos 2\varphi  - \gamma)^2  +  \frac{1}{2} \mu (1-\kappa) \beta ^2 \sin ^2 2\varphi \\
    + \frac{\gamma_D}{a^2} \ln \frac{1}{ 1-\frac{a^2}{b}|\beta_{,y}||\sin \varphi|
    }+ \tau_Y(\gamma) \beta \biggr]\,\mathrm{d} y,
\label{relaxedEnergy}
\end{multline}
with respect to $\beta $, provided the sign of $\dot{\beta}$ is positive during the loading course. For small up to moderate densities of excess dislocations the logarithmic terms in \eqref{SS_energyFinal} may be approximated by 
\begin{equation} \label{SS_log}
   \frac{\gamma_D}{a^2} \ln \frac{1}{ 1-\frac{a^2}{b}|\beta_{,y}||\sin \varphi|}
    \cong \frac{\gamma_D}{b}|\beta_{,y}||\sin \varphi|+\frac{1}{2}\frac{\gamma_D a^2}{b^2} \beta_{,y}^2 \sin^2 \varphi.
\end{equation}
Consequently the relaxed energy functional \eqref{relaxedEnergy} simplifies to
\begin{multline*}
I_d = w L \int_0^h \biggl[ \frac{1}{2} \mu \kappa \langle \beta \rangle ^2 \sin ^2  2\varphi +  \frac{1}{2}\mu (\langle \beta \rangle \cos 2\varphi  - \gamma)^2  +  \frac{1}{2} \mu (1-\kappa) \beta ^2 \sin ^2 2\varphi \\
    + \frac{\gamma_D}{b}|\beta_{,y}||\sin \varphi|+\frac{1}{2}\frac{\gamma_D a^2}{b^2} \beta_{,y}^2 \sin^2 \varphi+ \tau_Y(\gamma) \beta \biggr]\,\mathrm{d} y,
\end{multline*}
We shall deal further with  this functional only.

It is convenient to introduce the following dimensionless coordinates and quantities
\begin{equation*}
    E_d=\frac{I_d}{\mu w L h}, \, \eta=\frac{by}{a^2}, \, \bar{h}=\frac{bh}{a^2}, \, k=\frac{\gamma_D}{\mu a^2}=\frac{b^2}{4\pi (1-\nu)a^2},\, g(\gamma)=\frac{\tau _Y(\gamma)}{\mu },
\end{equation*}
in terms of which the above functional becomes
\begin{multline*}
E_d =\int_0^{\bar{h}} \biggl[ \frac{1}{2} \kappa \langle \beta \rangle ^2 \sin ^2  2\varphi +  \frac{1}{2} (\langle \beta \rangle \cos 2\varphi  - \gamma)^2  +  \frac{1}{2} (1-\kappa) \beta ^2 \sin ^2 2\varphi \\
    + k |\beta^\prime ||\sin \varphi|+\frac{1}{2}k \beta^{\prime 2} \sin^2 \varphi+ g(\gamma) \beta \biggr]\,\mathrm{d} \eta ,
\end{multline*}
where, now, the prime denotes the derivative with respect to $\eta$, and $\langle \beta \rangle =\frac{1}{\bar{h}}\int_0^{\bar{h}} \beta \,\mathrm{d}\eta $. Up to an unessential constant this functional can be reduced to
\begin{multline*}
E_d =\int_0^{\bar{h}} \biggl[ \frac{1}{2} \kappa \langle \beta \rangle ^2 \sin ^2  2\varphi +  \frac{1}{2} (\langle \beta \rangle \cos 2\varphi  - g_l(\gamma))^2  +  \frac{1}{2} (1-\kappa) \beta ^2 \sin ^2 2\varphi \\
    + k |\beta^\prime ||\sin \varphi|+\frac{1}{2}k \beta^{\prime 2} \sin^2 \varphi \biggr]\,\mathrm{d} \eta ,
\end{multline*}
which is similar to the functional studied in \citep{Le2008}. The only difference is that, instead of a constant $\gamma _l$ we have now
\begin{equation*}
g_l(\gamma)=\gamma -\frac{g(\gamma)}{\cos 2\varphi }.
\end{equation*}

Using the solution of the energy minimization problem found in \citep{Le2008} we determine the threshold value  $\gamma _{cr}$, when the non-uniform plastic slip begins, as the root of the equation
\begin{equation} \label{SS_energetic1}
    \gamma_{cr}= \frac{g(\gamma _{cr})}{\cos 2\varphi } + \frac{2 k a^2}{h b } \frac{\sin \varphi }{\cos 2 \varphi},
\end{equation}
provided $0 ^\circ < \varphi < 45^\circ$ (because only in this case the sign of $\dot{\beta}$ is positive). This equation demonstrates clearly the size effect for $\gamma_{cr}$. For $\gamma <\gamma_{cr}$ the plastic slip $\beta $ must vanish. For $\gamma >\gamma_{cr}$ the plastic slip reads
\begin{equation*}
\beta (\eta)=
  \begin{cases}
    \beta _1(\eta) & \text{for $\eta \in (0,l)$}, \\
    \beta _m & \text{for $\eta \in (l,\bar{h}-l)$}, \\
    \beta _1(1-\eta ) & \text{for $\eta \in (\bar{h}-l,\bar{h})$},
  \end{cases}
\end{equation*}
where 
\begin{equation*}
    \beta_1 = \beta_{1p} (1- \cosh \zeta \eta + \tanh \zeta l \sinh \zeta \eta), \quad 0 \leq \eta \leq l,
\end{equation*}
with $\zeta = 2 \sqrt{\frac{1-\kappa}{k}} \cos \varphi$ and
\begin{equation*}
    \beta _m=\beta_{1p} \left( 1-\frac{1}{\cosh \zeta l}\right).
\end{equation*}
Here $\beta_{1p}$ is given by
\begin{equation*}
    \beta_{1p}= \frac{g_l(\gamma) \cos 2 \varphi -  ( \cos^2 2 \varphi + \kappa \sin^2 2 \varphi) \langle  \beta \rangle}{(1-\kappa) \sin^2 2 \varphi},
\end{equation*}
while 
\begin{equation*}
    \langle \beta  \rangle= \frac{p(l)}{(1-\kappa)\bar{h} \sin^2 2\varphi + p(l) (\cos^2 2 \varphi+\kappa \sin^2 2\varphi) } g_l(\gamma) \cos 2\varphi ,
\end{equation*}
and
\begin{equation*}
p(l)=2\left(l-\frac{\tanh \zeta l}{\zeta }\right) +\left(1-\frac{1}{\cosh \zeta l} \right) (\bar{h}-2l).
\end{equation*}
Finally, the thickness of the boundary layers $l$ is the root of the transcendental equation
\begin{equation}
    f(l) \equiv 2 k \sin \varphi  -\frac{g_l(\gamma) \cos 2\varphi -\langle \beta  \rangle (\cos ^2 2\varphi +  \kappa \sin ^2 2\varphi )}{\cosh \zeta l}
    (\bar{h}-2l)=0 . \label{E:findL3}
\end{equation}
It is easy to extend this result to the case $45^\circ < \varphi < 90^\circ$ where sign$\, \dot{\beta}<0$ (see \citep{Le2008}).

\section{Numerical simulations}
\label{sec:4}

\begin{figure}[htb]
\centering \includegraphics[height=5cm]{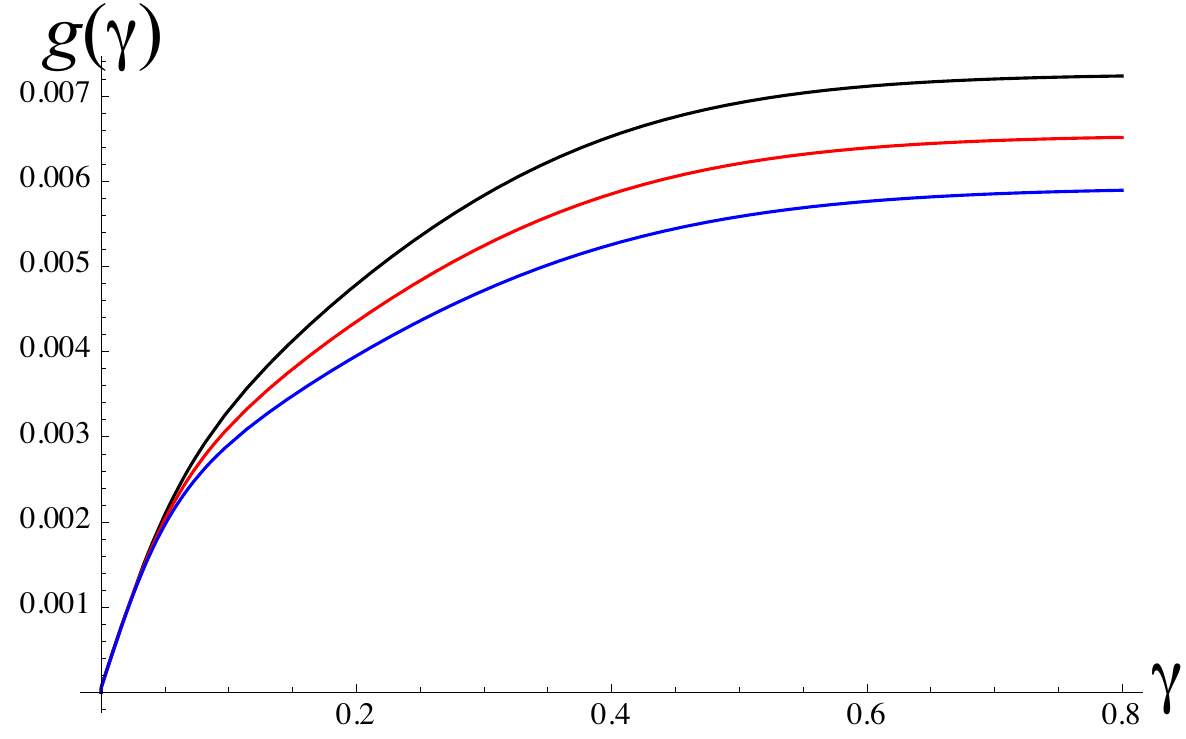} 
	\caption{Functions $g(\gamma )=\tau_Y(\gamma)/\mu$: (i) $\tilde{q}_0=10^{-12}$ (black),  (ii) $\tilde{q}_0=10^{-14}$ (red), (i) $\tilde{q}_0=10^{-16}$ (blue).}
	\label{fig:3}
\end{figure}

\begin{figure}[htb]
\centering \includegraphics[height=5cm]{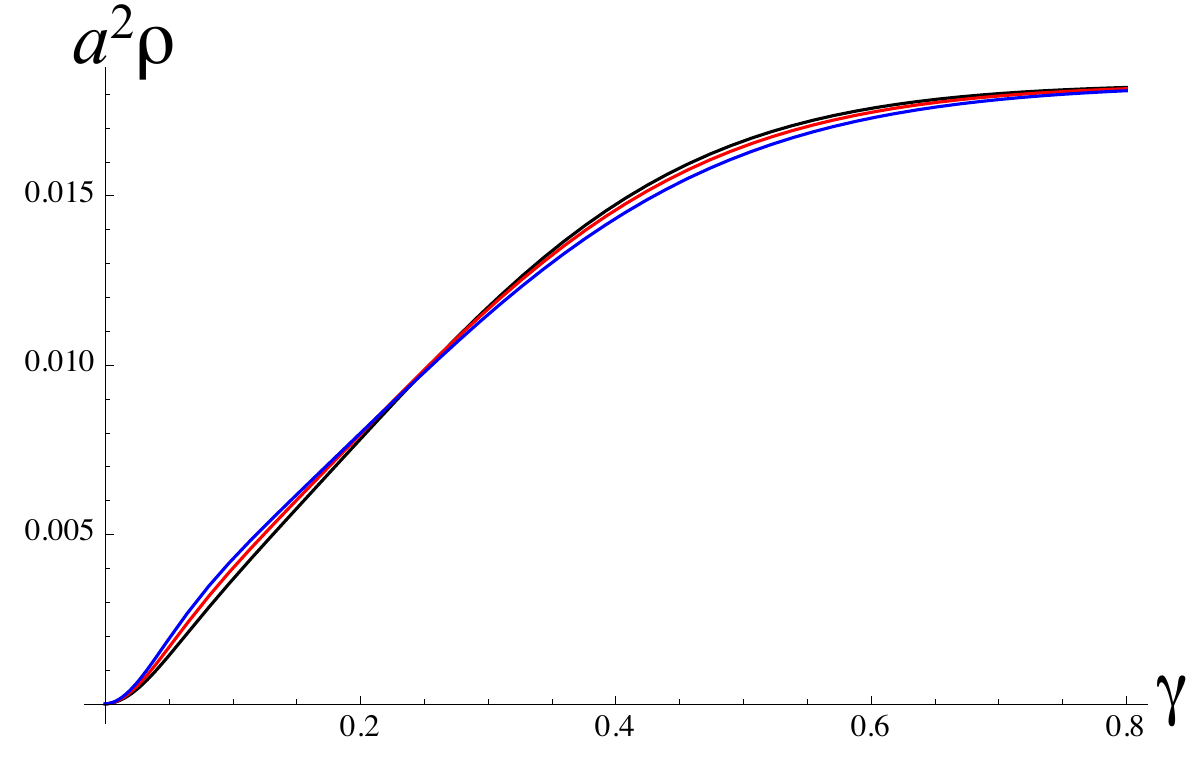} 
	\caption{Normalized total density of dislocations $\tilde{\rho }(\gamma )=a^2\rho $: (i) $\tilde{q}_0=10^{-12}$ (black),  (ii) $\tilde{q}_0=10^{-14}$ (red), (i) $\tilde{q}_0=10^{-16}$ (blue).}
	\label{fig:3a}
\end{figure}

Assume that the specimen is loaded with the constant shear strain rate $\dot{\gamma }$. As discussed in the previous Section, the first task is then to solve the system \eqref{eq:2.11}, with $\tau $ being replaced by $\tau _Y$ and $q_0=t_0 \dot{\gamma }\cos ^2 \varphi$. Since the shear strain rate $\dot{\gamma }$ is constant and only $g(\gamma )=\tau_Y(\gamma)/\mu$ is required for the next step, we choose $\gamma $ as the independent variables and rewrite this system of equations in terms of the following dimensionless quantities
\begin{equation*}
g(\gamma )=\frac{\tau _Y(\gamma )}{\mu },\quad \tilde{\rho }=a^2\rho , \quad \tilde{\chi }=\frac{\chi }{e_D}, \quad \tilde{\rho }^{ss}(\tilde{\chi})=e^{-1/\tilde{\chi}}.
\end{equation*}
\begin{figure}[htb]
\centering \includegraphics[height=5cm]{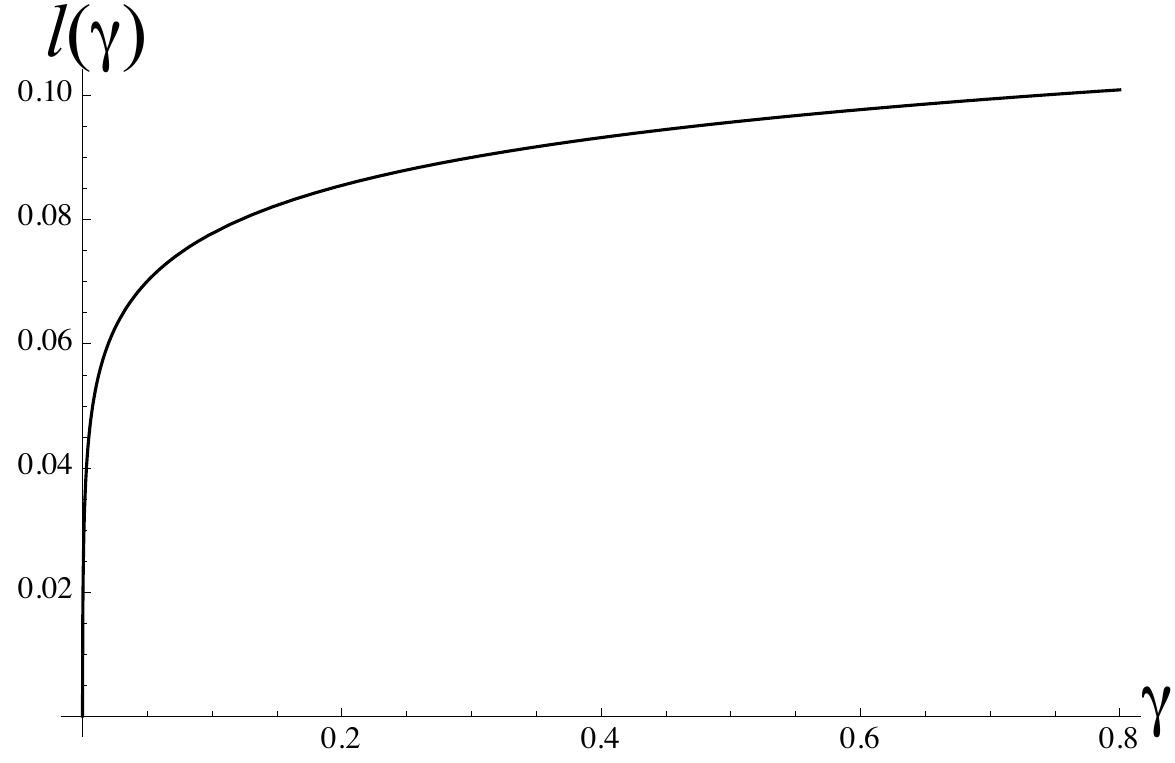} 	
	\caption{Relative thickness of the boundary layers $l(\gamma )$.}
	\label{fig:4}
\end{figure}
\begin{figure}[htb]
\centering \includegraphics[height=5cm]{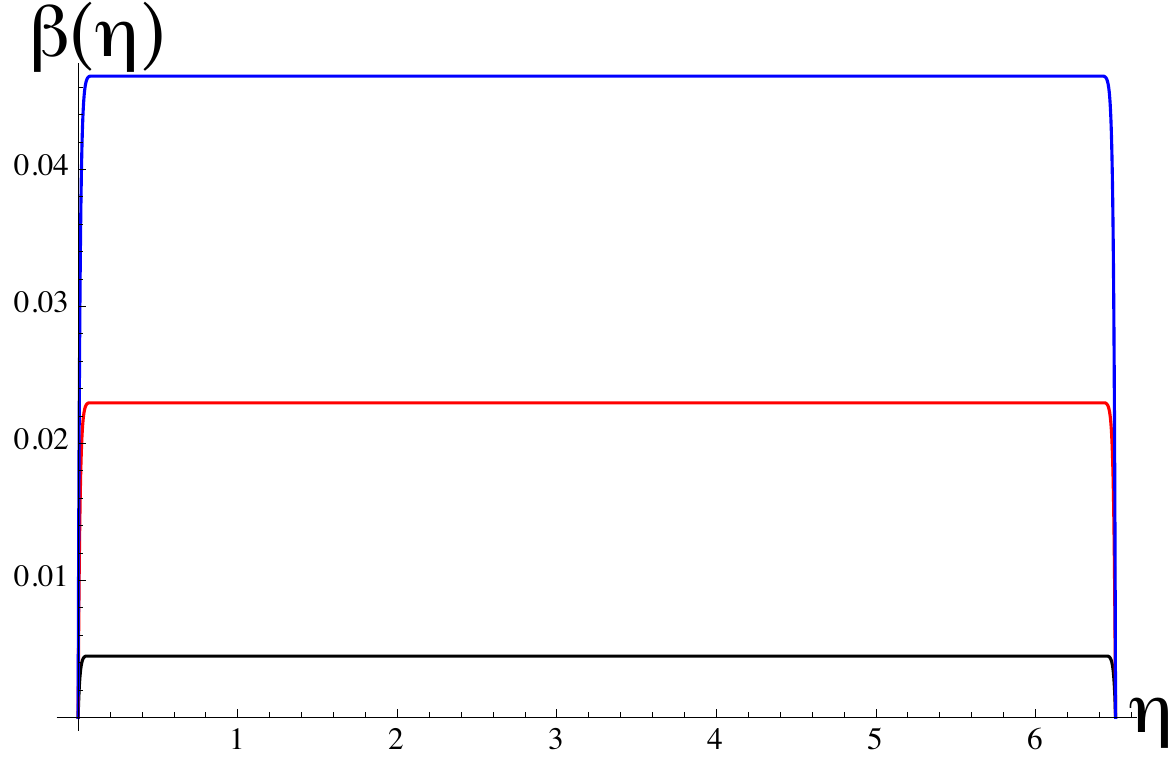} 	
	\caption{The plastic slip $\beta(\eta )$: (i) $\gamma =0.01$ (black), (ii) $\gamma =0.05$ (red), (iii) $\gamma =0.1$ (blue).}
	\label{fig:5}
\end{figure}

The system of ODEs becomes
\begin{align}
\frac{\mathrm{d}g}{\mathrm{d}\gamma } & = 1-\frac{\tilde{q}(g,\tilde{\rho })}{\tilde{q}_0} ,  \notag
\\
\frac{\mathrm{d}\tilde{\chi }}{\mathrm{d}\gamma } & = K g \frac{\tilde{q}(g,\tilde{\rho })}{\tilde{q}_0} \left[ 1-\frac{\tilde{\chi }}{\tilde{\chi }^{ss}(\tilde{q})} \right], \label{eq:4.2}
\\
\frac{\mathrm{d}\tilde{\rho }}{\mathrm{d}\gamma } & = \frac{K_\rho g }{\tilde{\nu }(\theta ,\tilde{\rho },\tilde{q}_0)^2}\frac{\tilde{q}(g,\tilde{\rho })}{\tilde{q}_0}\left[ 1-\frac{\tilde{\rho }}{\tilde{\rho }^{ss}(\tilde{\chi })} \right] . \notag
\end{align}

\begin{figure}[htb]
\centering \includegraphics[height=5cm]{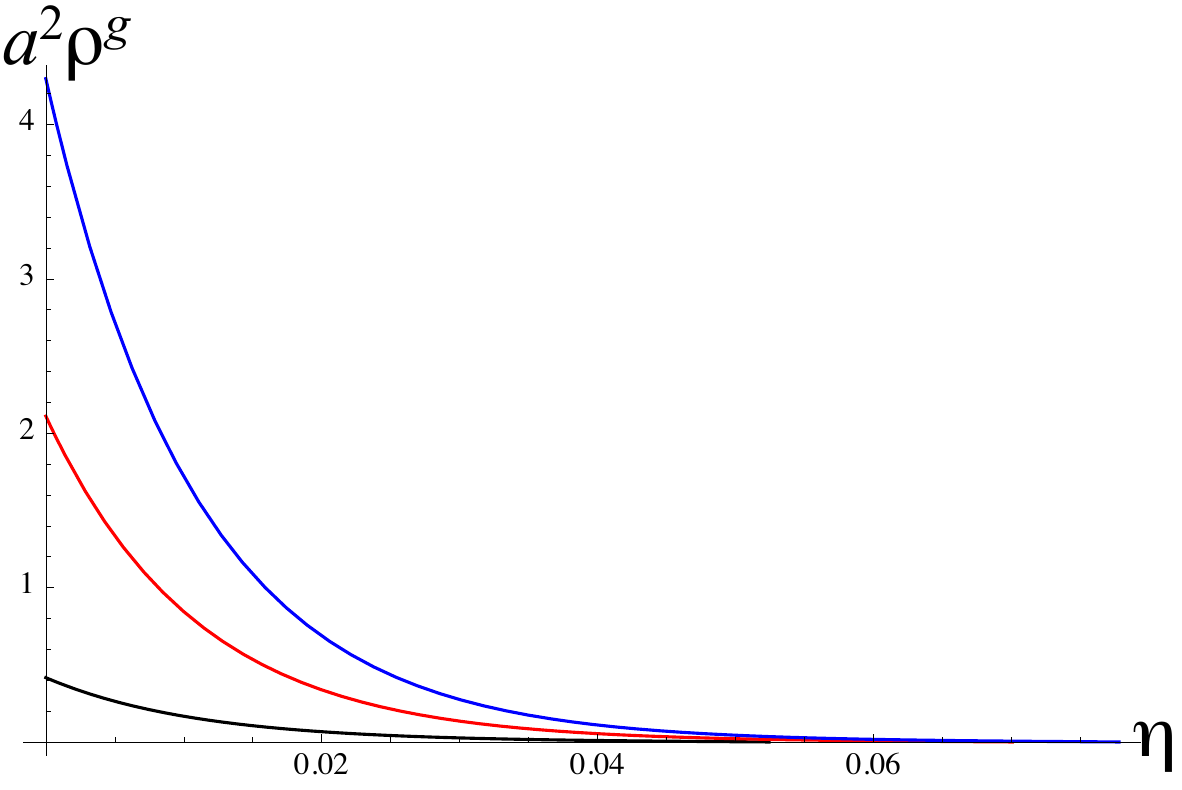} 
	\caption{The normalized density of excess dislocations $a^2\rho ^g=\beta ^\prime \sin \varphi $: (i) $\gamma =0.01$ (black), (ii) $\gamma =0.05$ (red), (iii) $\gamma =0.1$ (blue).}
	\label{fig:5a}
\end{figure}

Here $\tilde{q}_0=(a/b)\dot{\gamma }t_0\cos^2 \varphi $, $r=(b/a)\mu_T/\mu $, $K=\mathcal{K}\mu $, $K_\rho =\mathcal{K}_\rho \mu $ and
\begin{align*}
\tilde{q}(g,\tilde{\rho }) & = \sqrt{\tilde{\rho }}\exp \left[ -\frac{1}{\theta }e^{-g/(r\sqrt{\tilde{\rho }})}\right]  \\
\tilde{\nu }(\theta ,\tilde{\rho },\tilde{q}_0) & = \ln \left( \frac{1}{\theta }\right) - \ln \left[ \frac{1}{2}\ln \left( \frac{\tilde{\rho }}{\tilde{q}_0^2}\right) \right]  
\end{align*}
Let $T=298$ K. The parameters for copper at this room temperature are chosen as follows \citep{Langer2010} 
\begin{equation}
\label{eq:4.3}
r=0.0323, \quad \theta =0.0073, \quad K=350, \quad K_\rho =96.1,\quad \tilde{\chi }=0.25.
\end{equation}
We choose also the initial conditions 
\begin{equation}
\label{eq:4.4}
g(0)=0,\quad \tilde{\rho}(0)=10^{-6}, \quad \tilde{\chi}(0)=0.18.
\end{equation}
The plots of functions $g(\gamma )$ found by the numerical integration of \eqref{eq:4.2} for three different resolved shear strain rates are shown in Fig.~\ref{fig:3}. It can be seen that $g(\gamma )$ is rate-sensitive. Besides, it is also temperature-sensitive. Fig.~\ref{fig:3a} shows the evolution of the normalized total density of dislocations $a^2\rho $ versus $\gamma $ for the above shear strain rates.

\begin{figure}[htb]
\centering \includegraphics[height=5cm]{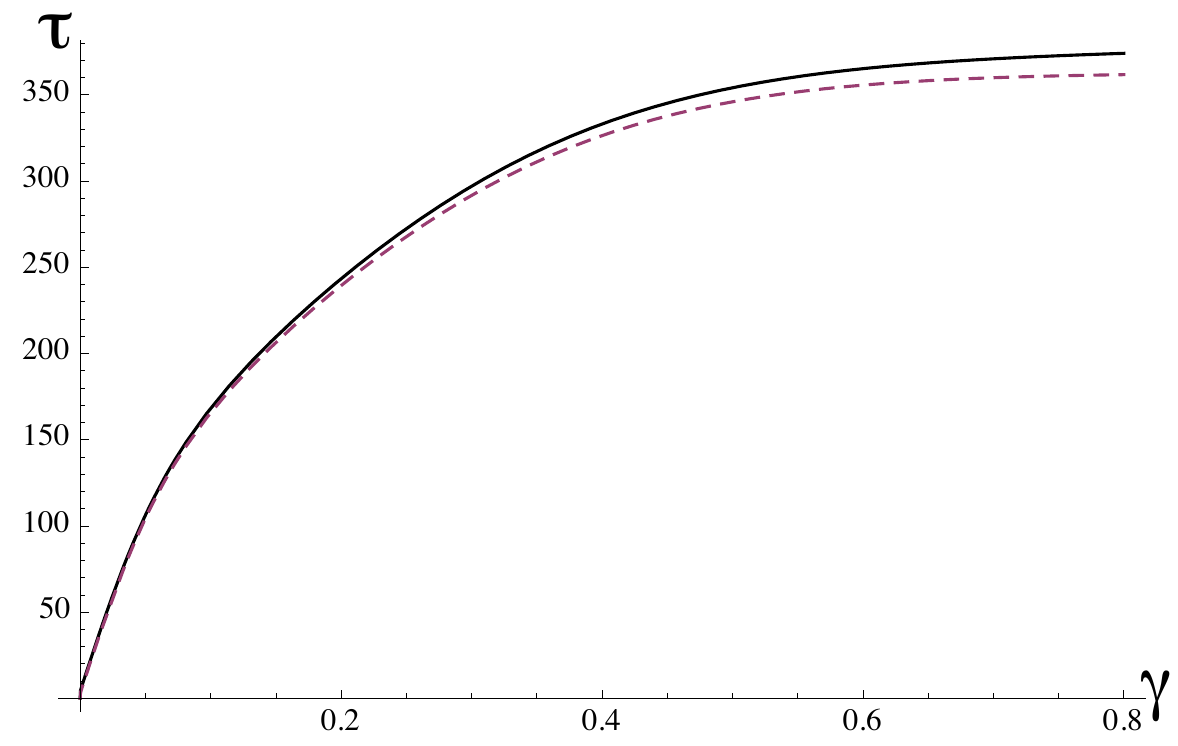} 
	\caption{Normalized resolved shear stress versus shear strain curve: (i) present theory (black), (ii) LBL-theory (dashed).}
	\label{fig:6}
\end{figure}

Having found $g(\gamma )$, we turn now to the determination of the plastic slip $\beta (\eta )$. In this problem let us fix $\tilde{q}_0=10^{-12}$ and choose the following parameters for copper
\begin{equation*}
\varphi =\pi /6,\quad b=0.255\, \text{nm}, \quad a=10\, \text{nm},\quad h=2.55\, \mu\text{m},\quad \nu=0.355.
\end{equation*}
First, the numerical solution of \eqref{SS_energetic1} gives the following threshold value: $\gamma _{cr}=0.000154$. For $\gamma >\gamma_{cr}$ Eq.~\eqref{E:findL3} has a single root that can be found numerically. The plot of function $l(\gamma)$ is shown in Fig.~\ref{fig:4}, from which it is seen that $l(\gamma )$ is a monotonically increasing function of $\gamma $. However, for the whole range of $\gamma \in (\gamma _{cr},0.8)$ the relative thickness of the boundary layers $l$ remains small compared to $\bar{h}=6.5025$. Next, we find the plastic slip as function of $\eta $ at three chosen values of $\gamma >\gamma _{cr}$. Their plots are shown in Fig.~\ref{fig:5}. We see that the plastic slip is constant in the middle of the specimen, and changes rapidly only in two thin boundary layers where positive and negative excess dislocations pile up against the grain boundaries. The number of excess dislocations increases with increasing shear strain. It is interesting to know the distribution of normalized density of excess dislocations $a^2\rho ^g=\beta ^\prime \sin \varphi $. Their distributions at three chosen values of $\gamma >\gamma _{cr}$ and in the lower boundary layers are shown in Fig.~\ref{fig:5a}. In the upper boundary layers the excess dislocations of opposite sign are symmetrically distributed. It turns out that, in this problem, the total number of excess dislocations
\begin{equation*}
N^g=2w\int_0^l \rho^g\, \mathrm{d}y=2w\frac{1}{b}\beta_m \sin \varphi 
\end{equation*}
is much smaller than the total number of dislocations $N=wh\rho $ for small strains. But for the strains of order 0.1 and larger the number of excess dislocations is comparable with the total number of dislocation, and the approximation \eqref{SS_log} may fail. Note that, due to the high gradient of the plastic slip in the boundary layers, the system \eqref{eq:2.5}, \eqref{eq:2.9}, and \eqref{eq:2.10} is no longer uncoupled there, so the more exact determination of the dislocation densities must be based on its solution that is possible only with the finite element method. Note also that, in other problems like the torsion of a bar or the bending of a beam, the percentage of excess dislocations becomes large even for small torque or bending moment.

\begin{figure}[htb]
\centering \includegraphics[height=5cm]{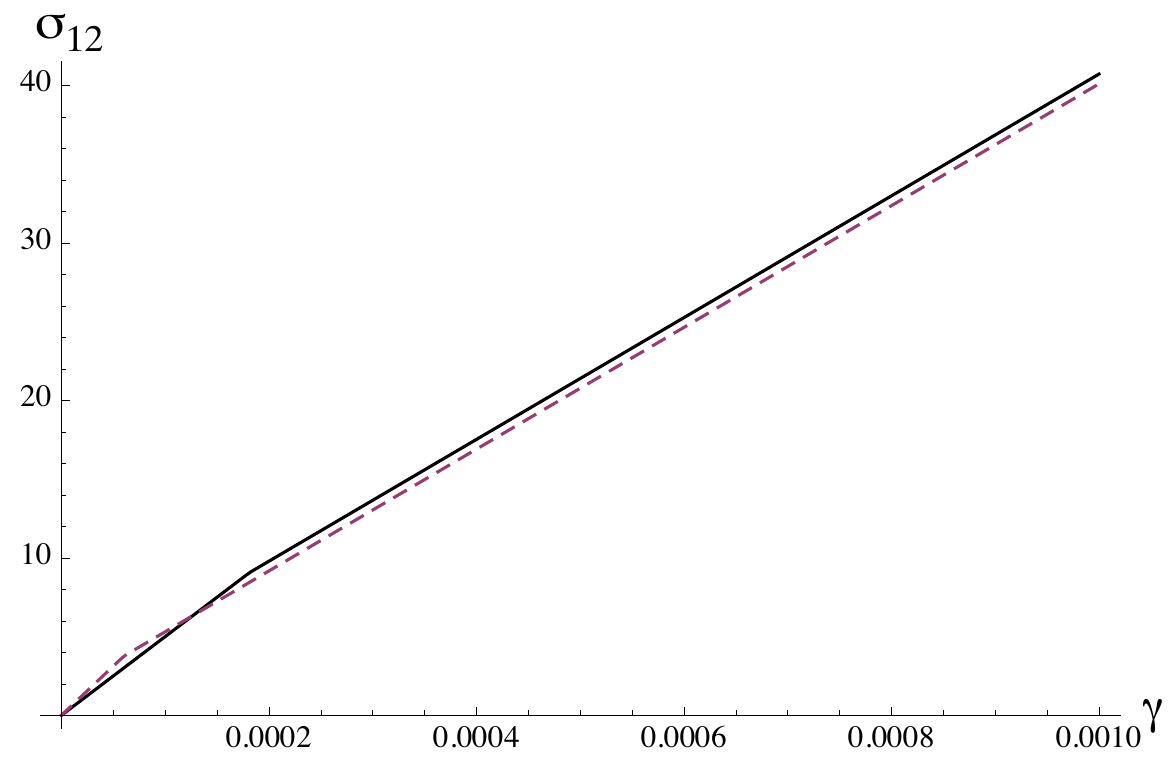} 
	\caption{Normalized shear stress $\sigma _{12}/\mu $ versus shear strain curve: (i) present theory (black), (ii) LBL-theory (dashed).}
	\label{fig:7}
\end{figure}

The resolved shear stress $\tau =\mu (\gamma \cos 2\varphi -\langle \beta  \rangle )$ (where $\mu=50\,$GPa) versus the shear strain curve is shown in Fig.~\ref{fig:6} together with the same curve $\mu g(\gamma )$ computed in accordance with the LBL-theory. We see that, in addition to the isotropic work-hardening caused by the redundant dislocations, there is a small kinematic work-hardening caused by the pile-up of excess dislocations. The difference due to this kinematic work-hardening becomes remarkable at large strains. Besides, it can be shown that this kinematic work-hardening depends on the thickness of the specimen, thus exhibiting the size effect (cf. \citep{Le2008}). Finally, we plot in Fig.~\ref{fig:7} the shear stress $\sigma_{12}= \mu (\gamma -\langle \beta  \rangle \cos 2\varphi )$ versus the shear strain curve. For comparison, the same shear stress computed from the LBL-theory $\sigma_{12}=\mu [\gamma (1-\cos^2 2\varphi )+g(\gamma )\cos 2\varphi ]$ is shown in the same Figure. Also here one can see the additional kinematic hardening due to the excess dislocations. At larger strains the two curves are nearly indistinguishable.

\section{Conclusion}\label{sec:5}

We have shown in this paper that the extension of the LBL-theory to non-uniform plastic deformations must necessarily take into account the excess dislocations whose density is expressed through the gradient of the plastic slip. This extension is shown to be consistent with the second law of thermodynamics. In the problem of plane strain constrained shear modeling a single crystal with grain boundaries, the excess dislocations are concentrated in thin boundary layers near the grain boundaries. The stress-strain curves exhibit both the isotropic hardening due to the redundant dislocations and kinematic hardening due to the pile-ups of excess dislocations against the grain boundaries which is size-dependent.

\bigskip
\noindent {\it Acknowledgement.}

The author thanks J.S. Langer for helpful discussions.

\end{document}